\documentclass{elsart}

\begin{document}

\hyphenation{smaller near studied present-ly hyper-geo-metric}
\hyphenation{respec-tively meso-scopic}

\begin{frontmatter}
\title{Nucleation of superconductivity in a mesoscopic loop
of finite width}
\author{Vital Bruyndoncx\thanksref{email}},
\author{Lieve Van Look},
\author{and Victor V. Moshchalkov}
\address{Katholieke Universiteit Leuven,
Laboratorium voor Vaste-Stoffysica en Magnetisme,
Celestijnenlaan 200 D, B-3001 Leuven, Belgium}
\thanks[email]{e-mail: Vital.Bruyndoncx@fys.kuleuven.ac.be}

\begin{abstract}

The normal/superconducting phase boundary $T_{c}$
has been calculated for mesoscopic loops, as a
function of an applied perpendicular magnetic field
H. While for thin-wire loops and filled disks the
$T_{c}(H)$ curves are well known, the intermediate
case, namely mesoscopic loops of finite wire width,
have been studied much less. The linearized first
Ginzburg-Landau equation is solved with the proper
normal/vacuum boundary conditions both at the
internal and at the external loop radius. For
thin-wire loops the $T_{c}(H)$ oscillations are
perfectly periodic, and the $T_{c}(H)$ background is
parabolic (this is the usual Little-Parks effect).
For loops of thicker wire width, there is a
crossover magnetic field above which $T_{c}(H)$
becomes quasi-linear, with the period identical to
the $T_{c}(H)$ of a filled disk (i.e. pseudoperiodic
oscillations). This dimensional transition is
similar to the 2D-3D transition for thin films in a
parallel field, where vortices start penetrating the
material as soon as the film thickness exceeds the
temperature dependent coherence length by a factor
1.8. For the presently studied loops, the crossover
point is controlled by a similar condition. In the
high field '3D' regime, a giant vortex state
establishes, where only a surface superconducting
sheath near the sample's outer radius is present.
\end{abstract}


\end{frontmatter}
\section{Introduction}
The nucleation of superconductivity in mesoscopic
samples has received a renewed interest after the
development of nanofabrication techniques, like
electron beam lithography. A superconductor is in
the mesoscopic regime when the sample size is
comparable to the superconducting coherence length
$\xi(T)$. In the framework of the Ginzburg-Landau
(GL) theory, $\xi(T)$ sets the length scale for
spatial variations of the modulus of the
superconducting order parameter $|\Psi|$. The
pioneering work on mesoscopic superconductors was
carried out already in 1962 by Little and
Parks~\cite{Litt62}, who measured the shift of the
critical temperature $T_{c}(H)$ of a (multiply
connected) thin-walled Sn microcylinder (a thin-wire
"loop") in an axial magnetic field $H$. The
$T_{c}(H)$ phase boundary showed a periodic
component, with the magnetic period corresponding to
the penetration of a superconducting flux quantum
$\Phi_0=h/2e$.

A few years later, Saint-James calculated the
$T_{c}(H)$ of a singly-connected
cylinder~\cite{SJ65cyl} (a mesoscopic "disk").
Taking into account the analogy with the situation
of a semi-infinite superconducting slab in contact
with vacuum~\cite{SJ63lin}, the critical field was
called $H_{c3}(T)$ in this case, since
superconductivity nucleates initially {\em near the
sample interface}. In the present paper, we will use
the notation $H_{c3}^{*}(T)$, for the nucleation
magnetic field.

The $T_{c}(H)$ phase boundary (or $H_{c3}^{*}(T)$)
of the disk shows, just like for the usual
Little-Parks~effect in a multiply connected sample
(loop), an oscillatory behaviour. When moving along
$T_{c}(H)$, superconductivity concentrates more and
more near the sample interface as $H$ grows. A giant
vortex state is formed: a "normal" core carries $L$
flux quanta, and the 'effective' loop radius
increases, resulting in a decrease of the magnetic
oscillation period. An experimental verification of
these predictions was carried out later on by
Buisson {\it et al.}~\cite{Buisson90} and by
Moshchalkov {\it et al.}~\cite{VVM95Nature}.

In the early paper from Saint-James and
de~Gennes~\cite{SJ63lin}, $H_{c3}^{*}(T)$ has been
calculated also for a {\em film exposed to a
parallel magnetic field}, where surface
superconductivity can grow along two
superconductor/vacuum interfaces. For low magnetic
fields, the two surface superconducting sheaths
overlap, and, as a result, $T_{c}$ versus $H$
becomes parabolic, which is characteristic for 2D
behaviour. When increasing the field, a crossover to
a linear $T_{c}(H)$ dependence (3D) occurs at $t
\approx 2 \, \xi(T)$, with $t$ the film thickness.
Shortly after, it was shown that vortices start to
nucleate in the film at this dimensional crossover
point ($t = 1.8 \,
\xi(T)$)~\cite{FinkSchultenscrossover}.

The goal of the present paper is to study the phase
boundary $T_{c}(H)$ of loops made of finite width
wires. In a Type-II material, superconductivity is
expected to be enhanced, with respect to the bulk
upper critical field
$H_{c2}$~:~$H_{c3}^{*}(T)>H_{c2}(T)$, both at the
external and the internal sample surfaces. As for a
film in a parallel field, a 2D-3D dimensional
crossover can be anticipated, since the loops may be
simply considered as a film, which is bent such that
its ends are joined together. We calculate, for the
first time, the phase boundary $T_{c}(H)$ as the
ground state solution of the linearized first
GL~equation with two superconductor/vacuum
interfaces.

\section{The linearized GL equation}
The linearized first GL equation to be solved in
order to find $T_{c}(H)$ is:
\begin{equation}
\frac{1}{2m^{\star}}(-i \hbar\vec{\nabla}-2 e
\vec{A})^{2}\Psi = |-\alpha| \, \Psi \; .
\label{glfree1}
\end{equation}
This equation is formally identical to the
Schr\"{o}dinger equation for a particle with a
charge $2 e$ in a magnetic field. At the onset of
superconductivity, the nonlinear GL term can be
omitted and the $z$-dependence disappears from the
equations and therefore an infinitely long cylinder
and a disk have identical $T_{c}(H)$ boundaries. It
is further assumed that $\mu_0 \vec{H}=rot\vec{A}$,
with $H$ the {\em applied magnetic field}. The
eigenenergies $|-\alpha|$ can be written as:
\begin{equation}
|- \alpha|=\frac{\hbar^{2}}{2 m^{\star} \:
\xi^{2}(T)}=\frac{\hbar^2}{2 m^{\star} \: \xi^{2}(0)}
\left( 1- \frac{T}{T_{c0}} \right) \; ,
\label{glalpha}
\end{equation}
$T_{c0}$ being the critical temperature in zero
magnetic field. From the energy eigenvalues of
Eq.~(\ref{glfree1}), the lowest Landau level
$|-\alpha_{LLL}(H)|$ is directly related to the
highest possible temperature $T_{c}(H)$, for which
superconductivity can exist.

For the loop geometries, we choose the cylindrical
coordinate system $(r,\varphi)$ and the gauge
$\vec{A}=(\mu_0 H r/2)\,
\vec{e}_{\varphi}$, where $\vec{e}_{\varphi}$ is the
tangential unit vector. The exact solution of the
Hamiltonian (Eq.~(\ref{glfree1})) in cylindrical
coordinates takes the following
form~\cite{Bezr9495,VBCross99}:
\begin{eqnarray}
\Psi(\Phi,\varphi)& = & e^{- \imath L \varphi}
\left( \frac{\Phi}{\Phi_0} \right)^{L/2}
\exp \left( - \frac{\Phi}{2 \, \Phi_0}
\right) K(-n,L+1,\Phi / \Phi_0) \label{psikummer} \\
K(a,c,y)& = & c_1 \, M(a,c,y)+ c_2 \, U(a,c,y) \, .
\nonumber
\end{eqnarray}
Here $\Phi=\mu_0 H \pi r^2$ is the applied magnetic
flux through a circle of radius $r$. The number $n$
determines the energy eigenvalue. Most generally,
the function $K(a,c,y)$ can be any linear
combination of the two confluent hypergeometric
functions (or Kummer functions) $M(a,c,y)$ and
$U(a,c,y)$~\cite{Abramowitz}, but the sample
topology puts a constraint on $c_1, \, c_2$, and
$n$, via the Neumann boundary condition:
\begin{equation}
\left. (- \imath \hbar \vec{\nabla} - 2 e \vec{A})
\Psi \right|_{\perp , b}=0 \; ,
\label{glbound}
\end{equation}
which the solutions $\Psi$ of Eq.~(\ref{glfree1})
have to fulfill at the sample interfaces $b$.

The eigenenergies of Eq.~(\ref{glfree1}) can be
written in the form:
\begin{equation}
\frac{r_o^2}{\xi^2 (T_{c})} =
\frac{r_o^2}{\xi^2(0)} \,
\left( 1- \frac{T_{c}(H)}{T_{c0}} \right)
= 4 \, \left( n+ \frac{1}{2}
\right) \, \frac{\Phi}{\Phi_0}
= \epsilon(H_{c3}^{*}) \, \frac{\Phi}{\Phi_0}  \, ,
\label{hcscaled}
\end{equation}
where $\Phi=\mu_0 H \pi r_o^2$ is arbitrarily
defined. The integer number $L$ is the phase
winding, or fluxoid quantum number. {\em The
parameter $n$ depends on $L$ and is not necessarily
an integer number}, as we shall see further on.

The bulk Landau levels are obtained when
substituting $n=0,1,2,\: \cdots$~ in
Eq.~(\ref{hcscaled}), meaning that the lowest level
$n=0$ corresponds to the bulk upper critical field
$\mu_0 H_{c2}(T)=\Phi_0/\left( 2 \pi
\xi^2(T)\right)$.

For a disk geometry~\cite{SJ65cyl,Buisson90}, we
have to take $c_2=0$ in Eq.~(\ref{psikummer}) in
order to avoid the divergency of $U(a,c,y
\rightarrow 0)=\infty$ at the origin. Selecting the
lowest Landau level at each value $\Phi$, one ends
up with a cusp-like $T_{c}(H)$ phase
boundary~\cite{SJ65cyl}, which is composed of values
$n<0$ in Eq.~(\ref{hcscaled}), thus leading to
$H_{c3}^{*}(T)>H_{c2}(T)$. A similar calculation was
performed for a single circular microhole in a plane
film ("antidot")~\cite{Bezr9495}, where $c_1=0$ in
Eq.~(\ref{psikummer}), since $M(a,c,y \rightarrow
\infty)=\infty$. Here as well, the lowest Landau
level consists of solutions with $n<0$. At each cusp
in $T_{c}(\Phi)$, the system makes a transition $L
\rightarrow L \pm 1$, i.e. a flux quantum enters or is
removed from the sample.

The loops we are currently studying have two
superconducting/vacuum interfaces, one at the outer
radius $r_o$, and one at the inner radius $r_i$.
Consequently, the boundary condition
(Eq.~(\ref{glbound})) has to be fulfilled at both
$r_o$, and $r_i$. As a result, we have a system of
two equations and two variables $n$ and $c_2$
($c_1=1$ is chosen), which we solved for different
values of $r_i/r_o$.

\section{Results}
Fig.~\ref{elevels} shows the Landau level scheme
(dashed lines) calculated from
Eqs.~(\ref{psikummer})-(\ref{hcscaled}), for a loop
with $r_i/r_o=0.5$. The applied magnetic flux
$\Phi=\mu_0 H \pi r_o^2$ is defined with respect to
the outer sample area. The $T_{c}(H)$ boundary is
composed of $\Psi$ solutions with a different phase
winding number $L$ and is drawn as a solid cusp-like
line in Fig.~\ref{elevels}. At $\Phi
\approx 0$, the state with $L=0$ is formed at $T_{c}(\Phi)$
and one by one, consecutive flux quanta $L$ enter
the loop as the magnetic field increases. For low
magnetic flux, the background depression of $T_{c}$
is {\em parabolic\/}, whereas at higher flux,
$T_{c}(\Phi)$ becomes {\em quasi-linear}, just like
for the case of a filled disk. The crossover point
from parabolic to quasi-linear appears at about
$\Phi
\approx 14 \, \Phi_0$.
\begin{figure}
\caption{Calculated energy level scheme (dashed lines) for a
superconducting loop with the ratio of inner to outer radius
$r_i/r_o=0.5$. The solid and dotted lines correspond to
$H_{c2}(T)$ and $H_{c3}(T)$, respectively.} \label{elevels}
\end{figure}

The solid and dotted straight lines in
Fig.~\ref{elevels} are the bulk upper critical field
$H_{c2}(T)$ and the surface critical field
$H_{c3}(T)$ for a semi-infinite slab, respectively.
In these units the slopes of the curves (see
Eq.~(\ref{hcscaled})) are $\epsilon=2$ for $H_{c2}$
(substitute $n=0$ in Eq.~(\ref{hcscaled})) and
$\epsilon=2/1.69$ for $H_{c3}$. The ratio
$\eta=\epsilon(H_{c2})/\epsilon(H_{c3})=1.69$
corresponds then to the enhancement factor
$H_{c3}(T)/H_{c2}(T)$ at a constant temperature. For
the loops we are studying here,
$\eta=\epsilon(H_{c2})/\epsilon(H_{c3}^{*})$ is
varying as a function of the magnetic field.

The energy levels below the $H_{c2}$ line (solid
straight line in Fig.~\ref{elevels}) could be found
by fixing a certain $L$, and finding the real
numbers $n<0$ numerically after inserting the
general solution (Eq.~(\ref{psikummer})) into the
boundary condition (Eq.~(\ref{glbound})). Note that
the lowest Landau level always has a lower energy
$|-\alpha(\Phi)|$ than for a semi-infinite
superconducting slab, which implies
$H_{c3}^{*}(T)>H_{c3}(T)=1.69 \, H_{c2}(T)$.

As mentioned earlier, in a thin film of thickness
$t$ in a parallel field $H$, a dimensional crossover
is found at $t=1.84 \, \xi(T)$. For low fields (high
$\xi$) $T_{c}(H)$ is parabolic (2D), and for higher
fields vortices start penetrating the film and
consequently $T_{c}(H)$ becomes linear
(3D)~\cite{FinkSchultenscrossover}. In
Fig.~\ref{elevels} the small arrow indicates the
point on the phase diagram $T_{c}(\Phi)$ where
$w=1.84 \, \xi(T)$. For the loops as well, the
dimensional transition shows up approximately at
this point, although the vortices are not
penetrating the sample area in the 3D regime.
Instead, the middle loop opening contains a coreless
'giant vortex' with an integer number of flux quanta
$L \, \Phi_0$.
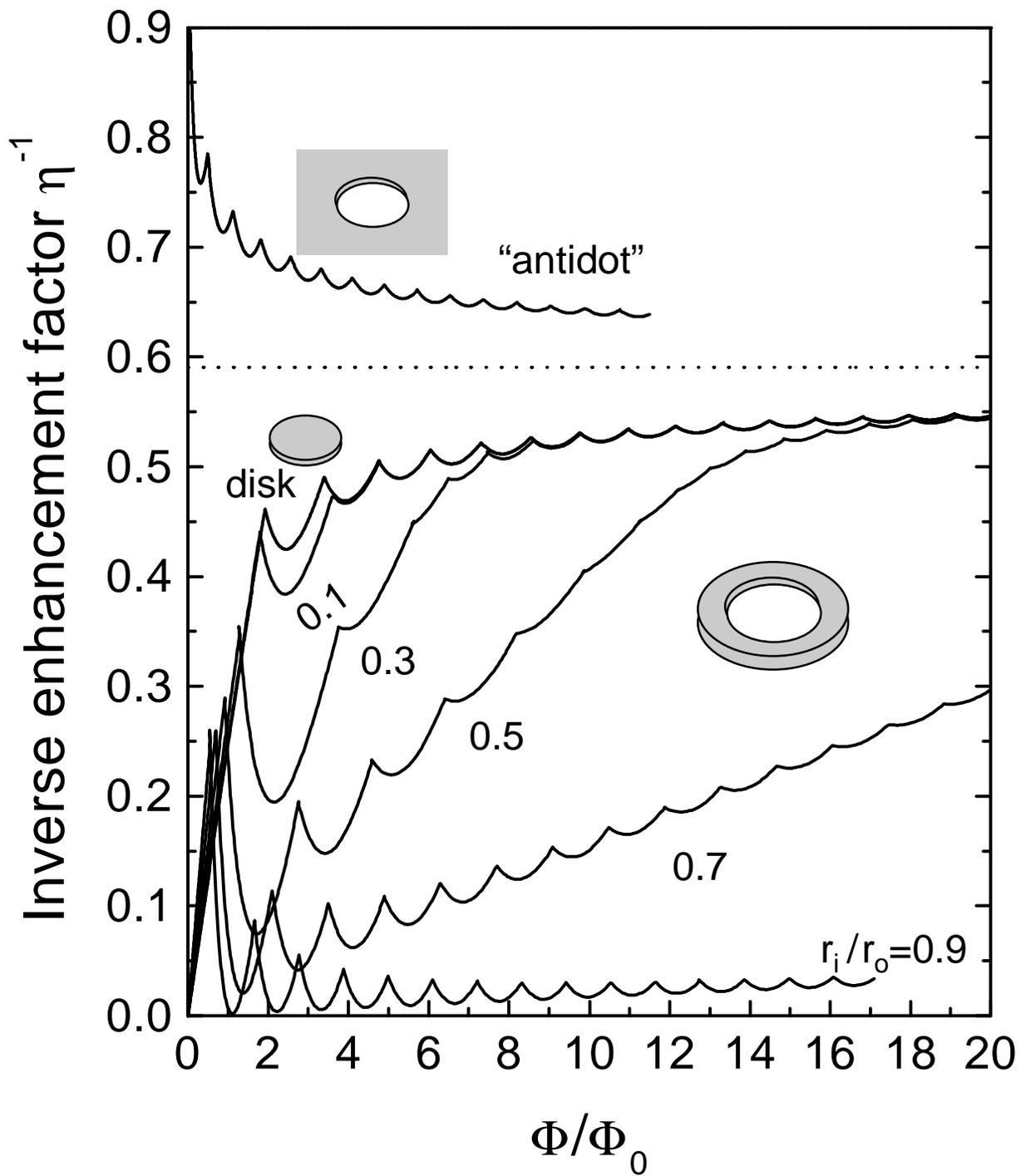
\begin{figure}
\caption{Inverse enhancement factor
$\eta^{-1}=\epsilon(H_{c3}^{*})/\epsilon(H_{c2})$
(see Eq.~(\ref{hcscaled})) for loops with different
aspect ratio, compared to the case of a disk and an
antidot. The horizontal dashed line at
$\eta^{-1}=0.59=1/1.69$ corresponds to
$H_{c3}(T)/H_{c2}(T)=1.69$ for a plane
superconductor/vacuum boundary~\cite{SJ63lin}.}
\label{enhancement}
\end{figure}

In order to compare the flux periodicity of
$T_{c}(\Phi)$, we have plotted, in
Fig.~\ref{enhancement}, the lowest energy levels of
Fig.~\ref{elevels} as
$\eta^{-1}=\epsilon(H_{c3}^{*})/\epsilon(H_{c2})$,
for loops with a different $r_i/r_o$. In this
representation, the dotted horizontal line at
$\eta^{-1}=0.59$ corresponds to the surface critical
field line $H_{c3}(T)$. The nucleation field of a
disk $H_{c3}^{*}(T)>1.69 \, H_{c2}(T)$, and for a
circular microhole in an infinite film
("antidot")~\cite{Bezr9495} $H_{c3}^{*}(T)<1.69 \,
H_{c2}(T)$. As $\Phi$ grows (the radius goes to
infinity) the $H_{c3}^{*}(T)$ of both the disk and
the antidot approaches the $H_{c3}(T)$ line.

For all the loops we study here, the presence of the
outer sample interface automatically implies that
$H_{c3}^{*}(T)>H_{c3}(T)$ is enhanced ($\eta>1.69$),
with respect to the case of a flat
superconductor-vacuum interface. For loops with a
small $r_i/r_o$, the $T_{c}(\Phi)$ boundary very
rapidly collapses with the $T_{c}(\Phi)$ of the dot
($\eta$ becomes the same). The presence of the
opening in the sample is not relevant for the giant
vortex formation in the high flux 3D regime. On the
contrary, in the low flux regime, the surface
sheaths along the two interfaces overlap, giving
rise to a different periodicity of $T_{c}(\Phi)$ and
to a parabolic background. This regime can be
described within the London limit\cite{Groff64},
since superconductivity nucleates almost uniformly
within the sample.

In summary, we have solved the linearized
GL~equation for loops of different wire width, with
Neumann boundary conditions at both the outer and
the inner loop radius. The critical fields
$H_{c3}^{*}(T)$ are always above the $H_{c3}(T)=1.69
\, H_{c2}(T)$. The ratio $H_{c3}^{*}(T)/H_{c2}(T)$ is
enhanced most strongly when the sample's
surface-to-volume ratio is the largest. The
$T_{c}(\Phi)$ behaviour can be split in two regimes:
for low flux, the background of $T_{c}$ is parabolic
(characteristic for 2D behaviour) and the
Little-Parks $T_{c}(\Phi)$ oscillations are
perfectly periodic. In the high flux regime, the
period of the $T_{c}(\Phi)$ oscillations is
decreasing with $\Phi$ and the background $T_{c}$
reduction is quasi-linear (3D regime). The 2D-3D
crossover between the two regimes, at a certain
applied flux $\Phi$, is similar to the dimensional
transition in thin films subjected to a parallel
field. As soon the 3D regime is reached, a giant
vortex state is created, where only a sheath close
to the sample's outer interface is superconducting.

\begin{ack}
The authors wish to thank H.~J.~Fink for stimulating
discussions. This work has been supported by the
Belgian IUAP, the Flemish GOA and FWO-programmes,
and by the ESF programme VORTEX.
\end{ack}

\end{document}